\documentclass[runningheads]{llncs}

\usepackage[show]{chato-notes}

\usepackage[T1]{fontenc}
\usepackage{graphicx}
\usepackage{enumitem} 

\usepackage{hyperref}
\usepackage{url}
\usepackage{multirow}
\usepackage{makecell}
\usepackage{calrsfs}
\usepackage{booktabs}
\usepackage{adjustbox}
\usepackage{caption}
\usepackage{subcaption}
\usepackage{tikz}
\usetikzlibrary{positioning}
\usepackage{mathtools}
\usepackage{amsmath}
\usepackage{soul}
\usepackage[most]{tcolorbox}
\usepackage[svgnames]{xcolor}

\usepackage{minted}

\usepackage{amssymb}
\usepackage{pifont}
\newcommand{\cmark}{\ding{51}}%

\usepackage{caption}
\setitemize{noitemsep}
\setlist{topsep=0pt, leftmargin=*}

\usepackage{orcidlink}

\newcommand{\uls}{\begin{itemize}[leftmargin=*]}
\newcommand{\ule}{\end{itemize}}
\newcommand{\ols}{\begin{enumerate}[leftmargin=*]}
\newcommand{\ole}{\end{enumerate}}
\newcommand{\li}{\item}

\newcommand{\para}[1]{\paragraph{\textnormal{\textbf{#1}}}}

\usepackage{siunitx}
\DeclareMathAlphabet{\pazocal}{OMS}{zplm}{m}{n}
\DeclareMathAlphabet{\pazobfcal}{OMS}{cmsy}{b}{n}

\newcommand{\nb}[3]{
    {\colorbox{#2}{\bfseries\sffamily\scriptsize\textcolor{white}{#1}}}
    {\textcolor{#2}{$\blacktriangleright$\textsf\small{#3}$\blacktriangleleft$}}}

\newcommand{\fzA}[1]{\textcolor{black}{#1}}

\newif\ifshowcomments
\showcommentsfalse 

\newcommand{\commentCM}[1]{
    \ifshowcomments{\nb{Comment (Craig):}{purple}{#1}}
    \fi
}
\newcommand{\commentDG}[1]{
    \ifshowcomments{\nb{Comment (Debasis):}{red}{#1}}
    \fi
}

\newcommand{\retriever}{\theta_{R}}
\newcommand{\topk}{\retriever(q)_k}
\newcommand{\llm}{\theta_{G}}
\newcommand{\pC}{Perp\textit{C}}
\newcommand{\pA}{Perp\textit{A}}

\commentCM{}
\commentDG{}

\usepackage{tikz}

\begin{document}
\title{Predicting Retrieval Utility and
Answer Quality in Retrieval-Augmented Generation}
\titlerunning{Predicting Retrieval Utility and Answer Quality in RAG}
%
\author{Fangzheng Tian\orcidlink{0009-0000-3282-0220} \and
Debasis Ganguly\orcidlink{0000-0003-0050-7138} \and
Craig Macdonald\orcidlink{0000-0003-3143-279X}}
\authorrunning{Tian et al.}
%
\institute{
University of Glasgow, Glasgow, UK
\\
\email{f.tian.1@research.gla.ac.uk},
\email{debasis.ganguly@glasgow.ac.uk},
\email{craig.macdonald@glasgow.ac.uk}
}
\maketitle              
\begin{abstract}
The quality of answers generated by large language models (LLMs) in retrieval-augmented generation (RAG) is largely influenced by the contextual information contained in the retrieved documents. A key challenge for improving RAG is to predict both the utility of retrieved documents---quantified as the performance gain from using context over generation without context---and the quality of the final answers in terms of correctness and relevance. In this paper, we define two prediction tasks within RAG. The first is retrieval performance prediction (RPP), which estimates the utility of retrieved documents. The second is generation performance prediction (GPP), which estimates the final answer quality. 
We hypothesise that \fzA{in RAG}, the topical relevance of retrieved documents correlates with their utility, suggesting that query performance prediction (QPP) approaches can be adapted for RPP and GPP. Beyond these retriever-centric signals, we argue that reader-centric features, such as \fzA{the LLM's perplexity of the retrieved context conditioned on the input query}, can further enhance prediction accuracy \fzA{for both RPP and GPP}. Finally, we propose that features reflecting query-agnostic document quality and readability can also provide useful signals to the predictions.
We train linear regression models with the above categories of predictors for both RPP and GPP. Experiments on the Natural Questions (NQ) dataset show that combining predictors from multiple feature categories yields the most accurate estimates of RAG performance.

\end{abstract}
\begin{figure}[tb]
\centering
\includegraphics[width=\columnwidth]{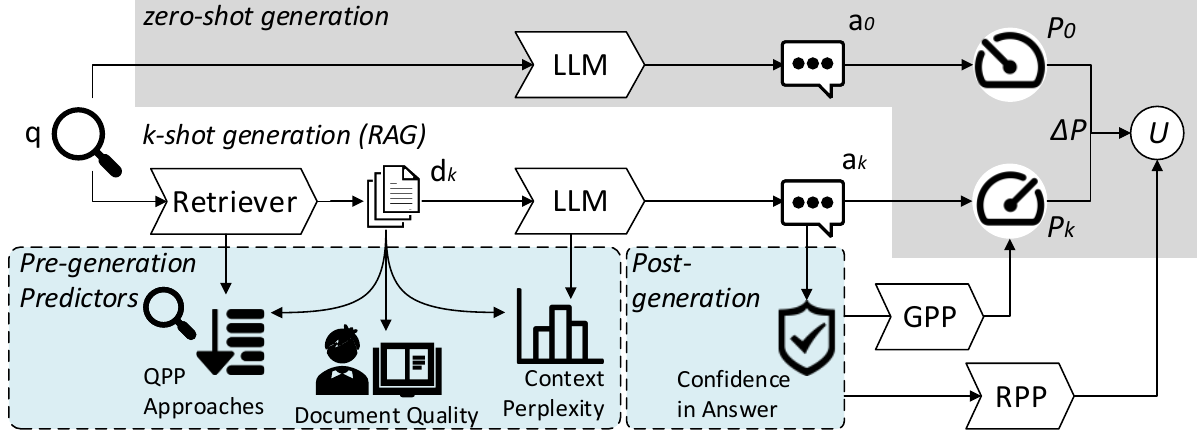}
\caption{
An illustration of how the proposed prediction tasks fit into a standard RAG workflow.
Retrieval Performance Prediction (RPP) estimates the gain in answer quality of $k$-shot relative to zero-shot (see Equation~\eqref{eq:utility}). Generation Performance Prediction (GPP) directly estimates the $k$-shot answer quality.
Both predictions rely on information extracted from the retrieved documents and the generated answer (as shown in the blue area at the bottom). The grey area denotes components not accessible to any predictors.}
\label{fig:flowchart}
\end{figure}

\section{Introduction}\label{s:introduction}
\looseness -1 Retrieval-Augmented Generation (RAG) is a framework that integrates the inherent parametric knowledge of large language models (LLMs) with retrieved documents to supplement knowledge~\cite{ragInKnowledgeIntensiveNLP}, and mitigate hallucinations~\cite{surveyOfHallucinations}.
Figure~\ref{fig:flowchart} shows how a list of top-$k$ documents retrieved for a query acts as a source for contextual generation via an LLM, a process often called ``$k$-shot generation''~\cite{ragReview}.

Although the standard RAG framework follows a static workflow that treats all queries uniformly~\cite{searchR1,ragInKnowledgeIntensiveNLP}, recent studies have shown the potential of adopting dynamic, query-specific workflows~\cite{oneSizeDoesntFitAll}. \fzA{In such dynamic workflows,} the number of retrieved documents or even the choice of ranking model may vary depending on the characteristics of the query~\cite{oldIRMeetRAG,FLARE,entityQueryChallengeForDenseRetr,adaptiveRAGforConv}.

It has been shown that predicting the retrieval quality in a standard IR task by query performance prediction (QPP) models~\cite{qppBERT-PL,denseQPP,QPP} leads to effective adaptive IR pipelines~\cite{DLinSelectiveRelevanceFeedback}. This motivates us to hypothesise that accurately predicting both the utility of retrieved documents as RAG context, where utility is defined as the answer quality gain of $k$-shot generation relative to zero-shot ($U$ in Figure~\ref{fig:linechart})~\cite{RelevanceAndUtility}, and the quality of the generated answers can further enhance RAG effectiveness. The primary objective of this work is therefore to investigate methods for estimating context utility and answer quality in RAG, while leaving the downstream integration of these estimators into adaptive RAG systems for future research.
To further clarify the two types of predictors examined in this paper, we consider two central questions when predicting RAG performance:
\uls
\li Do the retrieved documents improve answer quality compared to zero-shot generation? We call this task \textbf{retrieval performance prediction (RPP)}.
\li Do the retrieved contexts provide sufficient information for the LLM to generate correct answers that satisfy the query’s information need? We denote this as \textbf{generation performance prediction (GPP)}.
\ule
These two predictor components are illustrated in the bottom-right part of Figure~\ref{fig:flowchart}.
In addressing both RPP and GPP, we posit that the relevance of retrieved documents is a likely indicator of both context utility and answer quality, as conditional generation with relevant documents potentially leads to factually correct and relevant answers~\cite{RelevanceAndUtility}. However, since the ground-truth relevance is not known to a RAG system, the relevance estimated from existing QPP approaches may act as an effective proxy~\cite{qppInAgenticRAG}.  


However, different from standard IR, where the end user of the retrieved documents is a human, in RAG, the retrieved documents are consumed by an LLM~\cite{ragRelevanceLabel}. As a result, RPP and GPP should account not only for topical relevance but also for how the context interacts with the LLM~\cite{evalGenIR}.
This motivates us to explore indicators of quality other than relevance alone.

One of the factors on which the utility of a document depends is the perplexity, which captures how misaligned the retrieved context is with an LLM's own internal knowledge and semantics~\cite{infoTheoBasedPromptEng}.
To illustrate, in response to a sample query from the NQ dataset (as used in our experiments) ``\texttt{Who won the most MVP awards in the NBA?}'', a known relevant document's content -- \texttt{``the award a record six times. He is also the only player...}'' is topically related. However, as it omits the player’s name, when used as a RAG context, this document does not lead to the correct answer despite a zero-shot generation correctly outputting ``\texttt{Kareem Abdul-Jabbar}''.
\fzA{Because the player’s name is omitted, the LLM cannot reliably associate this relevant document with the target query, which manifests as higher conditional perplexity.
This example suggests that the perplexity of the retrieved context, reflecting the LLM’s uncertainty when conditioning on the input query, can serve as a useful predictive signal for both RPP and GPP.}

Beyond relevance and perplexity, \emph{query-agnostic} document quality characteristics such as readability and relevance priors may also influence the utility of retrieved contexts and the quality of generated answers. As an analogy, for human readers, overly complex text can limit usability~\cite{explPredOfComplexity}, and low-quality documents may fail to support any query~\cite{retrievability}.

In this work, we examine two variants of RPP and GPP: (a) \textbf{pre-generation} (\textbf{PreGen}) and (b) \textbf{post-generation} (\textbf{PostGen}). The PreGen predictor relies solely on information from the retriever, making it more efficient since it does not require the generator’s output. By contrast, the PostGen predictor can exploit additional signals, such as the perplexity of the generated answers~\cite{infoTheoBasedPromptEng}, thereby offering improved effectiveness at the expense of a modest reduction in efficiency.

\para{Our Contributions.} In summary, the main contributions of this paper are:
\uls
\li We apply existing QPP approaches for two novel prediction tasks in a RAG setting: retrieval performance prediction (RPP), and generation performance prediction (GPP).
\li In addition to relevance estimation via QPP, we further propose to leverage context perplexity and query-agnostic document quality measures to improve RPP and GPP.
\li We show that an ensemble of predictors learned via linear regression consistently outperforms the individual predictors across different retrieval models and context sizes.
\ule

\section{Related Work}\label{s:related_work}
\para{Predicting RAG Performance.}
In retrieval-augmented generation (RAG), standard IR metrics (e.g., nDCG) are insufficient proxies for utility because retrieved documents are consumed by an LLM rather than directly by human users~\cite{ragRelevanceLabel}. Prior work shows that context utility depends on multiple factors, including the context length~\cite{RelevanceAndUtility}, relevant document position~\cite{lostInTheMiddle}, prompt structure~\cite {powerOfNoise}, and knowledge conflicts between the retrieved document and the LLM’s parametric knowledge~\cite{knowledgeConflictOfLLM}.
\fzA{Existing approaches to predicting RAG answer quality mainly focus on uncertainty estimation. 
Answer-level semantic uncertainty is used to infer generation quality~\cite{measureRetrievalUtilityThruSemanticPerplexity,detectHallucinations,toolFormer,infoTheoBasedPromptEng}, but typically requires multiple sampled generations, limiting its practicality.
Token-level uncertainty has also been used to identify unreliable spans in the answer~\cite{FLARE,selfRAG,tokenLevelHarmonisationInRAG}.
In this line of work, \cite{predictingUtilityForTextCompletion} defines context utility as the reduction in answer uncertainty; however, this notion is disconnected from factuality and relevance, which are central to downstream evaluation.
Beyond uncertainty-based signals, some supervised methods predict whether retrieval is necessary by estimating query complexity~\cite{adaptiveRag,adaptiveRAGforConv}. More recently, \cite{utilityFocusedLLMAnnotationForRetrievalAndRAG,areLLMsGoodAtUtilityJudgements} directly apply LLMs to assess the utility of individual retrieved documents.}

\para{Query Performance Prediction (QPP).}
QPP methods estimate retrieval effectiveness by leveraging information from the top-retrieved documents and the query~\cite{QPP-benefits,QPP,QPP-goal}. 
Score-based approaches examine retrieval score distribution to assess the separation between relevant and non-relevant documents~\cite{Clarity,improveQPPbyStandardDeviation,NQC}, and the coherence of top-ranked results~\cite{coherenceQPP}.
Embedding-based methods exploit dense query–document representations to capture structural topology~\cite{denseQPP} and inter-document coherence~\cite{unsupervisedQPPforNeuralModelsWithPairwiseRankPreferences,aPairRatio}. 
Supervised models directly use the text of the query and the top-retrieved documents to predict IR metrics~\cite{bertQPP,deepQPP,qppBERT-PL,ContextRichQPP}.

\para{Reader-Oriented Document Evaluations.}

For human readers, document utility is often assessed through readability~\cite{FleschReadability}. 
Readability measurements typically rely on features such as sentence length, word difficulty, and syntactic complexity~\cite{ColemanReadability,KincaidReadability,SpacheReadability}. 
Beyond readability, the inherent quality of a document may also limit its usefulness, with some documents containing little or no valuable information~\cite{retrievability}. Recent approaches, such as QualT5~\cite{qualT5}, explicitly estimate query-agnostic document quality.
In RAG, whether these constraints affect LLMs as the new ``readers'' remains underexplored, highlighting potential differences in how humans and LLMs use retrieved text.

\para{Research Gap.}
\fzA{Prior work has not yet systematically studied context utility and answer quality prediction in RAG.
The closest effort, \cite{areLLMsGoodAtUtilityJudgements}, assesses the utility of individual retrieved documents rather than predicting the utility of RAG contexts concatenated from multiple documents. Moreover, it relies on LLM-based utility estimation and does not explore the use of existing retrieval analysis tools, such as QPP or document quality signals, in the RAG setting.}

\section{Predicting Retrieval Utility and Answer Quality}\label{s:task_description}
In this section, we formally define the tasks of RPP and GPP along with their prediction targets. We then introduce three types of predictors and propose combining them to improve prediction accuracy.

\subsection{Prediction Task Description}\label{ss:task_description}

\para{Preliminaries.}
Let $\retriever$ denote a retriever that maps a query $q$ to a ranked list of documents $\{d_1,\ldots, d_k\}$, and $\llm$ denote a generator that maps a prompt to an answer $a$. The $k$-shot answer is denoted as $a_k = \llm(q;\topk)$, where the prompt consists of the query $q$ and $\topk$, the latter denoting the top-$k$ retrieved documents obtained from $\retriever$ as context for $q$. Similarly, we can obtain a zero-shot answer, as $a_0 = \llm(q;\varnothing)$. If $k$-shot RAG is helpful, we would expect the generated answer $a_k$ to be enhanced compared to the zero-shot answer $a_0$. Indeed, \cite{RelevanceAndUtility} defined \emph{context utility} as the actual influence of retrieved documents on downstream answer quality. Specifically, \fzA{similar to \cite{RelevanceAndUtility},} we define the utility $U$ of the top-$k$ retrieved documents (when used as RAG context) as:
\begin{equation}
    U(\topk)=\pazocal{P}(\llm(q; \topk)) - \pazocal{P}(\llm(q; \varnothing)),
    \label{eq:utility}
\end{equation}
where $\pazocal{P}$ is a task-specific performance measure (e.g., F1 score for QA on NQ~\cite{evaluatingQAEval}).

\para{Retrieval Performance Prediction (RPP).}
The first prediction module in our RAG workflow, RPP, estimates the utility of retrieval results. The goal is to decide whether using the retrieved context is beneficial compared to zero-shot generation, without directly observing both $k$-shot and zero-shot performances as in Equation~\eqref{eq:utility}. Different from QPP in IR, which focuses on topical relevance, RPP targets context utility $U$ of the top-$k$ retrieved documents $\topk$, reflecting its support for LLM to answer the query $q$. Formally, an RPP model $\phi_{RPP}: q, \topk\mapsto\mathbb{R}$, where the output is an estimate of the utility $U$ of $\topk$. 
The prediction may be based on features of the query, the retrieved context, or (in post-generation settings) the generated answer. 
An accurate RPP model should assign higher predicted scores to the contexts that provide higher utility for answering $q$ in RAG. 

\para{Generation Performance Prediction (GPP).}

The second prediction module, GPP, differs from RPP in its prediction target. While RPP estimates the utility of retrieved documents relative to zero-shot generation, GPP directly predicts the quality of the generator’s final output under the current RAG configuration. A good GPP model can therefore indicate how reliable the produced answer will be, taking into account both the retrieved context and the LLM’s inherent knowledge.

Formally, the target of GPP is the answer quality $\pazocal{P}(a_k)$, computed against ground truth answers.
A GPP predictor is defined as $\phi_{GPP}: q, a_k \mapsto\mathbb{R}$, where the output is an estimate of the answer quality. 
An accurate GPP model should assign higher scores to answers that better satisfy the information need of the input query.
Equivalently, when ranking a set of generated answers, the GPP-induced ranking should be positively correlated with the ground-truth ranking by answer quality.
Similar to RPP, GPP can draw on multiple signals from both the retrieved context and the generated output, which we introduce next in Section~\ref{ss:predictors}.

\subsection{Information Sources for RPP and GPP}\label{ss:predictors}
Accurate predictions for a RAG system require analyses from multiple perspectives. Each analysis yields features related to retrieval utility and answer quality, which can be formulated as a predictor $\phi$. We group these predictors into three categories: (i) predicted relevance of the context, (ii) perplexity of the context or the generated answer, and (iii) intrinsic quality of the documents.

\para{Context Relevance Predictors.}

The first category of predictors predicts the relevance of the retrieved context, adapted from existing query performance prediction (QPP) approaches. Since the goal of a retriever is to return relevant documents, QPP methods analyse retrieval results from the retriever's perspective, and we therefore describe these as \textit{retriever-centric}. These predictors rely on signals available directly from the retrieval process, such as retrieval scores, dense embeddings, and document texts.

QPP encompasses a diverse set of methods, each targeting a different aspect of retrieval quality. Some estimate the upper bound of list relevance~\cite{aPairRatio}, \fzA{while others examine the distribution of the retrieval scores}~\cite{Clarity,NQC}. Still others focus on the coherence of retrieved documents~\cite{denseQPP}, which can also influence RAG performance. 
However, all QPP methods treat documents as independent units, making it difficult to capture cross-document dependencies and sentence-level interactions that emerge once the documents are concatenated and presented as a single context to an LLM in RAG~\cite{evalGenIR}.

\para{Perplexity-based predictors}

In contrast to QPP approaches, the second category of predictors is \textit{reader-centric}, analysing the RAG context and answer from the LLM’s perspective. These predictors leverage the token probabilities assigned by the model to estimate its internal certainty: given a query, how well the retrieved context aligns with the model’s expectations, and given both query and context, how confident it is in its generated answer~\cite{llmKnowsWhatItKnows}.
A widely-used measure of an LLM's confidence in its output is \emph{answer perplexity} (\pA{}), defined as the exponential of the mean negative log-probability of generated answer tokens~\cite{surveyUncertainQuantification}. Higher perplexity \pA{} is empirically associated with increased risk of hallucination~\cite{surveyOfHallucinations,tokenLevelHarmonisationInRAG}.

We extend this perplexity-based measurement to evaluate the retrieved context before generation. Given a query, if a retrieved context causes a lower perplexity as analysed by an LLM, it may indicate greater consistency with the model's expectations and inherent knowledge. To distinguish it from post-generation \pA, we refer to this measure as \emph{context perplexity} (\pC). 

\para{Document quality measurements.}

Unlike the previous two categories, the third category of predictors is \textit{query-agnostic}, providing estimations from the intrinsic complexity and quality of the retrieved documents. Readability measures assess text difficulty through factors such as word familiarity, sentence structure, and character statistics~\cite{FleschReadability}. More recently, supervised quality models estimate the inherent usefulness of a document independent of any query~\cite{qualT5}. These predictors complement retriever-centric and reader-centric signals by offering a baseline view of the context’s quality, regardless of query or model interpretation.

\subsection{Combining Multiple Predictions}\label{ss:combination}
Each potential predictor for RPP and GPP captures a different aspect of RAG performance. Hence by combining them, we can potentially perform a more comprehensive~\cite{qualityBiasedRankingOfWebDocuments} analysis of the retrieval results and generated answers.
By leveraging the complementary benefits of the predictors, such combinations can potentially lead to better estimates for both RPP and GPP,
as also observed in QPP applications~\cite{analysisVariationsInQPPEffectiveness}.
In particular, we construct an ensemble of the predictors by applying linear regression to capture the relationship between their outputs and the target values
---context utility for RPP and answer quality for GPP---on training queries. Formally,

\begin{equation}
\pazocal{L}(\Phi, Y; \mathbf{w}) = \sum_{q \in Q}(y(q)-\mathbf{w}\cdot \phi(q)
)^2,
\label{eq:loss_function}
\end{equation}
where $\mathbf{w} \in \mathbb{R}^n$ are the learnable weights of the predictor output $\phi(q) \in \mathbb{R}^n$ for query $q$, $Q$ is a set of training queries with ground-truth labels, and $y(q)$ denotes either the context utility for RPP, or the answer quality for GPP. \fzA{This loss function measures the squared error between the prediction $w\cdot \phi(q)$ and the target metric's ground truth $y(q)$. Minimising this loss corresponds to finding weights that produce predictions that are as close to the ground truth as possible, thereby achieving higher accuracy in RPP and GPP.}

\begin{figure*}[t]
\centering
\begin{adjustbox}{width=.85\columnwidth}
\begin{tcbraster}[%
    raster columns=1,
    raster equal height,
    raster column skip=1em,      
    nobeforeafter]
  \begin{tcolorbox}[%
      colback=white!15!white,
      colframe=black!60!white,
      ]
    You are an expert at answering questions based on your own knowledge and related context. Please answer this question based on the given context within 5 words. You should put your answer inside <answer> and </answer>.

    Question: \textit{Who won the most MVP awards in the NBA?}
    
    Doc 1: ....
    
    Doc 2: .... 
    
    Now start your answer. <answer> 
  \end{tcolorbox}
\end{tcbraster}
\end{adjustbox}
\caption{RAG prompt for generating answers for NQ datasets. The first part of an answer unit <answer> is put at the end to prompt the LLM to yield immediate answers.}
\label{fig:prompt-diagram}\vspace{-1em}
\end{figure*}

\section{Experimental Setup}\label{s:setup}

\para{Research Questions.}
To assess our framework's ability to predict context utility (RPP) and answer quality (GPP), we pose four research questions. RQ-1 evaluates existing QPP methods when applied directly to the two prediction tasks. RQs 2-4 progressively enrich the predictor set in the linear regression model described in Section~\ref{ss:combination}, trained for both tasks using the loss in Equation~\eqref{eq:loss_function}. Together, these RQs examine the contribution of different predictor categories and their combinations to RPP and GPP accuracy.  

\noindent -- \textbf{RQ-1}: How accurate are existing QPP approaches for RPP and GPP?

\noindent -- \textbf{RQ-2}: Does adding reader-centric context perplexity (\pC{}) improve prediction accuracy over QPP alone?  

\noindent -- \textbf{RQ-3}: Do query-agnostic document quality and readability metrics provide additional predictive value for RPP and GPP?  

\noindent -- \textbf{RQ-4}: Does integrating pre-generation predictors with post-generation answer perplexity (\pA{}) improve prediction accuracy? 

\para{Datasets.}

We evaluate the prediction accuracy of our proposed approaches on Natural Question~\cite{NQ} (NQ), a widely-used open-domain QA dataset. Answer quality is evaluated by F1 score against the provided golden answer~\cite{evaluatingQAEval}. Context documents for NQ queries are retrieved from a snapshot of English Wikipedia from 2018. The accuracy of predictors in RPP and GPP is tested on the NQ test set (3610 queries). Prediction accuracy is measured by Spearman's $\rho$ correlation between the predicted scores and the ground-truth targets, specifically context utility \fzA{\eqref{eq:utility}} for RPP, and answer quality (F1 Score) for GPP.

\para{RAG Configurations.}

To assess the applicability of our predictors for RPP and GPP, we experiment with three retrieval configurations\footnote{We perform all indexing and retrieval using the PyTerrier framework~\cite{pyterrier}.}:
\uls
\li \textbf{BM25}~\cite{BM25}: A lexical retriever based on term frequency, inverse document frequency, and document length normalisation.

\li \textbf{BM25 $\gg$ MonoT5}~\cite{monot5}: A retrieve-and-rerank pipeline, where the top-100 BM25 results are re-ranked using MonoT5, a cross-encoder re-ranker.

\li \textbf{E5}~\cite{E5}: A BERT-based bi-encoder trained with contrastive learning, leveraging large batch sizes and in-batch negatives. 
\ule

\noindent The above three rankers yield different overall answer quality on NQ in our experiments, with average F1 scores of 0.3284, 0.4019, and 0.4798, respectively (top-2 retrieved documents as context). 

We vary the RAG context size with $k \in \{2,3,5,7,10\}$. For answer generation, we employ an 8-bit quantised, instruction-tuned version of Llama-3-8B\footnote{\href{https://huggingface.co/QuantFactory/Meta-Llama-3-8B-GGUF}{QuantFactory/Meta-Llama-3-8B-GGUF}}~\cite{llama3}, guided by the prompt template shown in Figure~\ref{fig:prompt-diagram} that instructs the model to produce concise and factually correct answers.

\subsection{Investigated Predictors}

\para{QPP Approaches.}  We experiment with existing QPP approaches:

\uls
\li\textbf{NQC}~\cite{NQC}, which analyses the variance of retrieval scores of the top-retrieved documents, integrating with query term frequency.

\li \textbf{MaxScore}~\cite{aPairRatio}, which uses the score of the top-ranked document as an estimation of the retrieved context's upper bound on relevance.

\li \textbf{Dense-QPP}~\cite{denseQPP}, which computes the volume of the minimal hypercube covering the embeddings of the query and its top-$k$ documents ($k$ equals the context size). E5 is used to obtain the embeddings~\cite{E5}.

\li \textbf{A-Pair-Ratio}~\cite{aPairRatio}, which measures coherence by comparing pairwise similarities between the top-5 and bottom-5 documents from the top-100.

\li \textbf{BERT-QPP}~\cite{bertQPP}, a cross-encoder trained to predict MRR from the text of a query and the corresponding top-retrieved document.
\ule

\para{Perplexity-Based Predictors.}
As introduced in Section~\ref{ss:predictors}, we use context perplexity (\pC{}) and answer perplexity (\pA{}) as reader-centric predictors. Both predictors are derived from the model’s next-token probabilities provided by Llama-3-8B. In our experiments, \pC{} serves as a pre-generation predictor, while \pA{} is used as an additional post-generation predictor for both RPP and GPP. 
For implementation, we compute \pA{} and \pC{} as the exponentials of the average log-probabilities (without negation), so that higher values correspond positively with answer quality and context utility.

\para{Document Quality Measures.} 
To capture both document- and context-level signals, we compute the maximum, minimum, and average score across individual documents, as well as a direct score for the concatenated context, using the following document readability and quality measures:
\uls
\li \textbf{Traditional Readability Metrics}, including Dale-Chall~\cite{DaleChallReadability} and Spache~\cite{SpacheReadability}, which rely on word familiarity and lexical difficulty; Flesch–Kincaid~\cite{FleschReadability} and Gunning Fog~\cite{GunningReadability}, which analyse sentence structure complexity; Coleman–Liau~\cite{ColemanReadability}, which uses character-based counts as proxies for reading effort.

\li \textbf{QualT5}~\cite{qualT5}: a supervised T5-based model trained to estimate document quality in a query-agnostic manner, designed to assess the intrinsic usefulness of documents for downstream retrieval tasks.
\ule

We use linear regression (Equation~\eqref{eq:loss_function}) to combine 2 or more predictors, trained on the NQ dev set (8757 queries), using context utility ($U$) and answer quality ($\pazocal{P}$) as targets for RPP or GPP, respectively.

\begin{table*}[t]
\centering
\caption{Prediction accuracy of single QPP predictors in RPP (left) and GPP (right). Results are reported on the NQ test set using the top-2 retrieved documents as RAG context ($k$=2). The best method for each retrieval configuration is bold-faced.}\label{table:system_summarisation}
\begin{adjustbox}{width=0.65\textwidth}

\begin{tabular}{l rrr rrr}
\toprule
& \multicolumn{3}{c}{RPP} & \multicolumn{3}{c}{GPP} \\
\cmidrule(lr){2-4} \cmidrule(lr){5-7}
QPP Method & \multicolumn{1}{c}{BM25} & \multicolumn{1}{c}{MonoT5} & \multicolumn{1}{c}{E5} & \multicolumn{1}{c}{BM25} & \multicolumn{1}{c}{MonoT5} & \multicolumn{1}{c}{E5} \\
\midrule
NQC & 0.0694 & -0.0643 & 0.0994 & 0.0610 & -0.1288 & 0.1210 \\
MaxScore & 0.1485 & 0.1625 & 0.1288 & 0.1609 & 0.2218 & 0.2344 \\
DenseQPP & 0.1739 & \textbf{0.2027} & 0.1025 & \textbf{0.2166} & \textbf{0.2992} & 0.1922 \\
A-Pair-Ratio & 0.1565 & 0.0136 & \textbf{0.1913} & 0.1353 & -0.0483 & \textbf{0.2470} \\
BERT-QPP  & \textbf{0.1898} & 0.1474 & 0.1352 & 0.2079 & 0.1641 & 0.1630 \\
\bottomrule
\end{tabular}

\end{adjustbox}
\label{table:qpp_table}
\end{table*}
\begin{table*}[t]
\centering
\small
\caption{
Prediction accuracy of combining various predictor groups using linear regression, for $k=2$ retrieved documents.
The best scores for the PreGen and the PostGen settings are bold-faced separately.
A ``$^\dagger$'' indicates statistical significance (Fisher’s $z$, 95\% confidence) of a PreGen predictor over a QPP-only setting, whereas a ``$^\ddagger$'' indicates statistical significance of a PostGen predictor over a PreGen setting.
}\label{table:ablation_table}
\begin{adjustbox}{width=.9\textwidth}
\begin{tabular}{lccccc SS SS SS}
\toprule
& \multicolumn{5}{c}{Features ($\phi$)} & \multicolumn{3}{c}{RPP} & \multicolumn{3}{c}{GPP} \\
\cmidrule(r){2-6} \cmidrule(r){7-9} \cmidrule(r){10-12}
Type & \pA & QPP & \pC & Read & Qual & \multicolumn{1}{c}{BM25} & \multicolumn{1}{c}{MonoT5} & \multicolumn{1}{c}{E5} & \multicolumn{1}{c}{BM25} & \multicolumn{1}{c}{MonoT5} & \multicolumn{1}{c}{E5} \\
\midrule
\multirow{4}{*}{\makecell{Pre\\Gen}} & & \cmark &  &  &  & 0.2146 & 0.2008 & \textbf{0.2155} & 0.2454 & 0.2861 & 0.2952 \\
& & \cmark & \cmark &  &  & 0.2365 & 0.2001 & 0.2110 & 0.2821$^{\dagger}$ & 0.3040 & \textbf{0.3094} \\
 & & \cmark & \cmark & \cmark &  &
 \textbf{0.2419} & 0.2136 & 0.2101 & \textbf{0.2948}$^{\dagger}$ & 0.3071 & 0.3075 \\
 & & \cmark & \cmark & \cmark & \cmark & 0.2380 & \textbf{0.2350}$^{\dagger}$ & 0.2093 & 0.2945$^{\dagger}$ & \textbf{0.3253}$^{\dagger}$ & 0.3088 \\
\midrule
\multirow{5}{*}{\makecell{Post\\Gen}} & \cmark &  &  &  &  & 0.1468 & 0.1351 & 0.1470 & 0.3037 & 0.2579 & 0.2873 \\
& \cmark & \cmark &  &  &  & 0.2387 & 0.2283 & 0.2513$^{\ddagger}$ & 0.3507$^{\ddagger}$ & 0.3616$^{\ddagger}$ & 0.3915$^{\ddagger}$ \\
& \cmark & \cmark & \cmark &  &  & 0.2534 & 0.2280 & 0.2474$^{\ddagger}$ & 0.3653$^{\ddagger}$ & 0.3726$^{\ddagger}$  & \textbf{0.3988}$^{\ddagger}$ \\
& \cmark & \cmark & \cmark & \cmark &  &
\textbf{0.2573} & 0.2404 & \textbf{0.2476}$^{\ddagger}$ & \textbf{0.3729}$^{\ddagger}$ & 0.3737$^{\ddagger}$ & 0.3977$^{\ddagger}$ \\
& \cmark & \cmark & \cmark & \cmark & \cmark & 0.2539 & \textbf{0.2563} & 0.2472$^{\ddagger}$ & 0.3727$^{\ddagger}$ & \textbf{0.3856}$^{\ddagger}$ & 0.3974$^{\ddagger}$ \\
\bottomrule
\end{tabular}
\end{adjustbox}
\label{table:combined_table}
\end{table*}

\section{Results}\label{s:results}
We first report the results using the top-2 retrieved documents as context in Section~\ref{ss:observations} to answer the research questions. 
Then, we discuss how prediction accuracy varies with context size in Section~\ref{ss:along_k}.

\subsection{Main Observations}\label{ss:observations}
\para{RQ1: Usefulness of QPP approaches for RPP and GPP.}
Table~\ref{table:qpp_table} reports the results of existing QPP approaches for RPP and GPP with each column corresponding to a particular ranker.
Overall, QPP estimates are positively correlated with both context utility (RPP) and answer quality (GPP), confirming that topical relevance remains an informative signal for predicting RAG performance, consistent with prior findings~\cite{RelevanceAndUtility}. 
From Table~\ref{table:qpp_table}, we observe that QPP methods generally achieve higher correlations in GPP than in RPP, suggesting that answer quality is easier to approximate than predicting utility because it likely depends not only on the retrieved context’s relevance but also on how that information interacts with the LLM’s inherent knowledge~\cite{knowledgeConflictOfLLM}.
The most effective QPP predictor varies with the retriever model used -- consistent with earlier findings that QPP accuracy is ranker-dependent~\cite{analysisVariationsInQPPEffectiveness}. For BM25 and MonoT5, DenseQPP leads to the best GPP results indicating that the density of the document embeddings is a useful indicator of the answer quality.
However, the effective performance of MaxScore with MonoT5 and A-pair ratio on E5 indicates that no QPP approach works consistently the best across all rankers.

\textbf{To conclude for RQ-1}, existing QPP approaches exhibit different applicability to RPP and GPP in the RAG pipeline. While some approaches achieve good accuracy, their accuracy is not consistent across retrieval configurations.

\para{RQ-2: Effectiveness of combining QPP approaches with context perplexity (\pC{}).}
Table~\ref{table:combined_table} reports results when combining multiple predictors using linear regression as introduced in Section~\ref{ss:combination}. The first row in the upper part shows that aggregating multiple QPP predictors generally outperforms the best individual QPP method in Table~\ref{table:qpp_table}, especially for BM25 and E5. 

The second row in Table~\ref{table:combined_table} shows the results of combining QPP and reader-centric \pC{}. Adding \pC{} consistently improves prediction accuracy in GPP across all retrievers (e.g., the significant improvement from 0.2454 to 0.2821 with BM25), confirming our hypothesis that context perplexity influences the LLM to digest the context and produce high-quality answers. For RPP, however, gains in prediction accuracy appear only with BM25. This suggests that \pC{} is more discriminative for weaker retrievers, where the perplexity of retrieved contexts varies more sharply between low- and high-utility cases.

\textbf{To conclude for RQ-2}, combining \pC{} with QPP predictors yields more significant gains in accuracy with weaker retrievers and is especially effective for predicting answer quality rather than predicting context utility.

\para{RQ-3: Effectiveness of leveraging query-agnostic predictors.} 
The last two rows in the upper part of Table~\ref{table:combined_table} report results when adding query-agnostic readability (Read.) and document quality (Qual.) predictors to the predictor combination.
Compared with QPP and \pC{}, we observe small but consistent gains, mostly for BM25 and MonoT5. For BM25, adding readability alone already achieves the best pre-generation results. 
These findings suggest that document quality metrics, originally designed for human readers, offer only limited additional value for RAG prediction, where the reader is an LLM.

\textbf{To conclude for RQ-3}, document quality and readability predictors can slightly improve RPP and GPP accuracy when combined with QPP and \pC{}. 

\para{RQ-4: Effectiveness of Leveraging Post-Generation Prediction.}
The lower part of Table~\ref{table:combined_table} reports results for answer perplexity (\pA) and its integration with pre-generation predictors. Stand-alone \pA{} shows a moderate positive correlation with answer quality in GPP (Spearman’s $\rho$ at 0.25–0.30), but weaker correlations with retrieval utility in RPP, since utility $U$ also depends on how retrieved documents interact with the LLM’s inherent knowledge, a factor not fully reflected in the LLM's confidence of the answer~\cite{surveyUncertainQuantification}.

When \pA{} is combined with pre-generation signals, prediction accuracy improves for both RPP and GPP compared to either \pA{} or the pre-generation signals alone. For instance, in GPP with E5, combining \pA{} with QPP achieves a correlation of 0.3915, compared to 0.2873 for \pA{} alone and 0.2952 for QPP. These improvements are consistently significant under Fisher’s $z$-test, indicating that the model’s self-confidence, captured by \pA{}, complements the pre-generation signals from the context to yield stronger predictions.
Nonetheless, this benefit comes at the cost of waiting for the full answer to be generated.

\textbf{To conclude for RQ-4}, incorporating post-generation \pA{} consistently improves RPP and GPP accuracy over pre-generation predictors.

\begin{figure}[tb]
\centering
\includegraphics[width=\columnwidth]{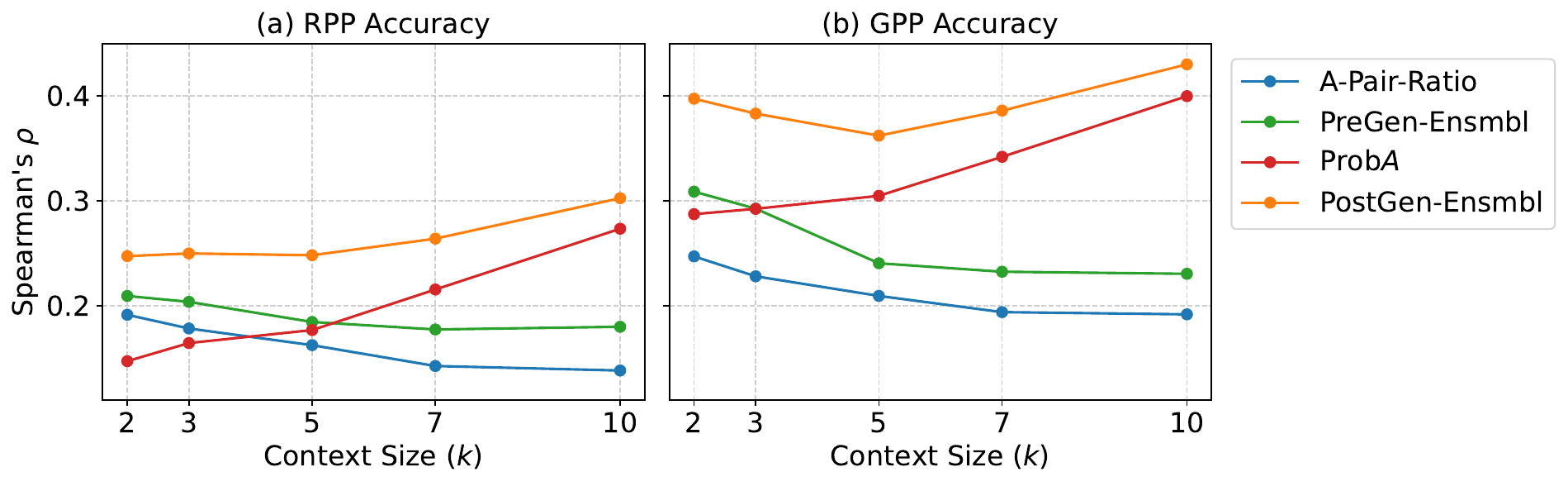}
\caption{Prediction accuracy with varying context size from 2 to 10. Sub-graph (a) shows results for RPP, and (b) for GPP. The retriever is E5 for both settings.}
\label{fig:linechart}
\end{figure}

\subsection{Prediction Accuracy for Varying Context Size ($k$)}\label{ss:along_k}
As the size of the RAG context (in terms of the number of documents in the context) changes, both answer quality and context utility may vary~\cite{RelevanceAndUtility}, which in turn affects the prediction accuracy of RPP and GPP.
Figure~\ref{fig:linechart} shows the results for the E5 retriever across context sizes from 2 to 10.
We compare: (1) the best-performing pre-generation method (A-Pair-Ratio, see Table~\ref{table:qpp_table}); (2) the combination of all pre-generation predictors (PreGen-Ensmbl); (3) the post-generation predictor \pA{}; and (4) the full combination of all pre- and post-generation predictors (PostGen-Ensmbl).

At each context size, ensembles consistently outperform individual predictors: PreGen-Ensmbl is always more accurate than the best single pre-generation method, and PostGen-Ensmbl yields the best overall performance. This aligns with our earlier findings in RQs, where combining predictors produces more accurate predictions than using them in isolation.
Overall, the accuracy of pre-generation predictors declines as $k$ grows for both RPP and GPP, reflecting the increasing difficulty of analysing long concatenated contexts. Although combining all pre-generation predictors improves accuracy compared to the best individual predictor, it cannot counteract the downward trend. 
The gains achieved by combining pre-generation predictors become larger as $k$ in RPP, suggesting that reader-centric and query-agnostic signals help capture inter-document relations missed by QPP alone, especially for long RAG contexts.

By contrast, the post-generation predictor \pA{} improves steadily as the context size increases. When incorporated with pre-generation predictors, they yield the best overall prediction performance. However, as $k$ increases, the contribution of pre-generation signals in the combination diminishes, as shown by the convergence of the curves of \pA{} and PostGen-Ensmbl in Figure~\ref{fig:linechart}.

\section{Conclusions}\label{s:conclusion}
In this paper, we propose two prediction tasks for RAG: \textbf{RPP}, targeting context utility, and \textbf{GPP}, targeting answer quality.
Our experiments show that a range of different information sources --- relevance estimations from QPP models, perplexity of retrieved and generated content, readability and relevance priors --- is useful for RPP and GPP.
An ensemble approach learned via linear regression consistently outperforms the individual predictors across different rankers and context sizes.
We also find that predicting utility (RPP) is inherently harder than predicting answer quality (GPP), since utility depends not only on relevance but also on how retrieved context interacts with an LLM’s inherent knowledge.
\fzA{While our results shed light on prediction-driven analysis for RAG, effectively leveraging such predictions in per-query context optimisation requires improving predictor precision. In addition, evaluating RPP and GPP across a broader range of datasets and LLMs is important for future work, as context utility may vary with task and model choice.}

\begin{credits}

\subsubsection{\discintname}
There is no competing interest.
\end{credits}
%
%
%
%
\bibliographystyle{splncs04}

\end{document}